\documentclass[aps,prl,twocolumn,superscriptaddress,a4paper]{revtex4-1}

\usepackage{fancyhdr} 

\usepackage{hyperref} 

\usepackage{verbatim} 
\usepackage{amsmath} 
\usepackage{amsthm} 
\usepackage{amssymb}	
\usepackage{graphicx} 
\usepackage[dvips,letterpaper,margin=0.75in,bottom=0.5in]{geometry}
 
\renewcommand{\d}[2]{\frac{d #1}{d #2}} 
\newcommand{\dd}[2]{\frac{d^2 #1}{d #2^2}} 
\newcommand{\pd}[2]{\frac{\partial #1}{\partial #2}} 
 
\let\baraccent=\= 
\renewcommand{\=}[1]{\stackrel{#1}{=}} 

\theoremstyle{definition}

\theoremstyle{remark}

\newcommand{\eps}{\varepsilon} 

\providecommand{\e}[1]{\ensuremath{\times 10^{#1}}} 

\begin{document}

\title{Intense Ion Beam Neutralization Using Underdense Background Plasma}


\author{William Berdanier}
\thanks{The authors are grateful to the Heavy Ion Fusion Science Virtual National Laboratory (HIFS-VNL) scientists, engineers and technologists for their experimental data. This work was supported by the National Undergraduate Fellowship Program and by the Department of Energy. Work at the Lawrence Berkeley Lab was supported by the Office of Science of the US Department of Energy under contract no. DE-AC02Ð05CH11231.}
\affiliation{Department of Physics, The University of Texas at Austin, Austin, Texas 78712}
\affiliation{Princeton Plasma Physics Laboratory, Princeton University, Princeton, New Jersey 08543}
\author{Prabir K. Roy}
\affiliation{Lawrence Berkeley National Laboratory, Berkeley, CA 94720}
\affiliation{Department of Nuclear Engineering \& Radiological Sciences, University of Michigan, Ann Arbor, MI 48109 (current address)}
\author{Igor Kaganovich}
\affiliation{Princeton Plasma Physics Laboratory, Princeton University, Princeton, New Jersey 08543}


\date{\today}

\begin{abstract}
Producing a dense background plasma for neutralization purposes is experimentally difficult and requires a large amount of energy. We show that even an underdense background plasma with a small relative density can achieve high neutralization of intense ion beam pulses. Using particle-in-cell simulations, we show that if the total plasma electron charge is not comparable to the beam charge, electron emitters are necessary for effective neutralization, but are not needed if the plasma volume is large. Several plasma densities are investigated, including the case of electron emitters without plasma, which does not effectively neutralize the beam. Over 95\% neutralization is achieved for even very underdense background plasma. We compare with an analytic model of neutralization and find close agreement with the particle-in-cell simulations. Further, we show experimental data from the NDCX-II group that verifies the result that underdense plasma can neutralize intense heavy ion beams effectively.
\end{abstract}

\maketitle

The space-charge neutralization and focusing of intense charged particle beams by background plasma forms the basis for a variety of applications for astrophysics\cite{alfven_motion_1939,bennett_magnetically_1934,medvedev_long-time_2005}, atomic physics\cite{smith_separation_1947}, high-energy accelerators and colliders\cite{chen_acceleration_1985,govil_observation_1999,joshi_development_2007}, basic physics phenomena\cite{soloshenko_physics_1996} and inertial confinement fusion, in particular, fast ignition\cite{roth_fast_2001,campbell_simulation_2005,mason_heating_2006} and heavy ion fusion\cite{r._c._davidson_physics_2001,roy_drift_2005,sefkow_optimized_2007,yu_heavy_2005}. One of the modern approaches to ion beam compression for heavy ion fusion applications is to propagate the ion beam through a dense background plasma, which charge neutralizes the ion bunch. Heavy-ion fusion requires that the ion beam be compressed and focused onto the target, which implodes upon impact; recent progress has been made in the focusing of neutralized beams by strong solenoidal magnetic fields\cite{kaganovich_charge_2007,dorf_enhanced_2009,qin_centroid_2010,qin_class_2013}. Neutralization facilitates compression of the bunch against strong space-charge forces, and is thus a key ingredient in any heavy ion fusion scheme. The focus of this paper will be to show that underdense plasma can provide a high degree of charge neutralization.

The required degree of space charge neutralization can be estimated from the beam envelope equation:

\begin{equation}
\dd{r_b}{z} = \frac{Q}{r_b} + \frac{\eps^2}{r_b^3},
\label{eq:envelope}
\end{equation}

where $Q=2\pi e^2 Z_b^2 n_b r_b^2/\gamma_b^3 M V_b^2$ ($Z_b$ is the charge state of the beam ions, $n_b$ is the beam density, $r_b$ is the beam radius, $\gamma_b$ is the relativistic factor of the beam, $M$ is the beam ion mass and $V_b$ is the beam velocity). For heavy ion fusion applications, the self-electric potential due to the space charge of the ion beam pulse is between approximately 100 V at the chamber entry to as much as 10 kV at the chamber exit\cite{roy_drift_2005,yu_heavy_2005}. This is much larger than the temperature of the ion beam, which is set by the ion source emitter and is of order 0.1 eV\cite{roy_drift_2005,yu_heavy_2005}, so we can neglect the emittance term, $\eps^2/r_b^3$, in the beam envelope equation\cite{m._reiser_theory_1994,r._c._davidson_physics_2001}. Integrating Eq. \ref{eq:envelope}, we obtain:

\begin{equation}
\left(\d{r_b}{z}\right)^2 = {r'_i}^2 + 2 Q \ln\left(\frac{r_b}{r_i}\right).
\end{equation}

For ballistic focusing, the beam space charge has to be neutralized enough so that the beam convergence angle $r'=dr_b/dz$ is not affected by the self-fields of the beam pulse during the drift. Thus, from Eq. 2 it follows that the degree of charge neutralization, $f$, should satisfy:

\begin{equation}
2(1-f)Q\ln\left(\frac{r_i}{r_f}\right) < {r'_i}^2.
\end{equation}

For the National Drift Compression Experiment (NDCX) at Lawrence Berkeley National Laboratory, $Q$ is of order $10^{-3}$, the beam radius in the extraction region of the ion beam source is $r_i \sim 2.5 \ \mathrm{cm}$, the compressed beam radius is $r_f \sim 1 \ \mathrm{mm}$, and the initial beam convergence angle $r'_i$ is of order $10^{-2}$\cite{roy_drift_2005,yu_heavy_2005,friedman_beam_2010}. Thus, the degree of neutralization should be better than $(1-f)<10^{-2}$. Many different schemes have been investigated to achieve this high degree of neutralization, and only propagation through background plasma has been shown to be viable\cite{kaganovich_physics_2010}. Past studies have investigated the use of dense plasma ($n_p\ge n_b$) for neutralization; however, producing dense plasma is experimentally difficult and requires a large amount of energy. Thus, it is highly advantageous to investigate the neutralization capabilities of lower background plasma densities. 

The purpose of this Letter is to demonstrate that a high level of neutralization can be obtained from propagating the pulse through very underdense plasma, where $n_p\ll n_b$, so long as the plasma has more space charge than the beam pulse. If the plasma has less space charge than the beam pulse, an electron source on the chamber walls is necessary. We consider two possible electron sources on the walls of a chamber filled with a preformed, quiescent, underdense plasma: (1) electron emitters, a computational boundary condition that produces particles when a normal electric field is present, or (2) a region of very dense plasma on the walls (here, $n_{\mathrm{wall}} = 10 n_b$). The first scheme is computationally less intensive, since the boundary produces particles only whenever a normal electric field is present, and is thus more easily modeled. This simulates the physics of electrons being pulled from some other source, such as a grid of hot wires. The second scheme is more experimentally realistic, as one of the commonly used ways to produce a quiescent, underdense plasma in a chamber is to place a plasma source on the walls. This leaves a region of dense plasma near the source\cite{efthimion_long_2009,gilson_ferroelectric_2014}. 
 
\begin{figure} [h]
\caption{ (a) Self-electric field of un-neutralized beam in vacuum. (b) Self-electric field of beam for the electron-emitters-only neutralization scheme. The electric field is  $\sim40$ kV/cm, so the hot electrons from the wall do not effectively neutralize the beam pulse. (c) Beam self-electric field from neutralization from volumetric plasma of density $n_p = n_b/50$. The total plasma space charge is 0.49 that of the beam, and neutralization is poor. (d) Beam self-electric field after neutralization by volumetric plasma of density $n_p = n_b/5$. The total plasma space charge is 5 times that of the beam, and the self-electric field reduction is 98\%. Note the strong electric fields near the edge of the chamber.} 
(a)\includegraphics[width=0.43\columnwidth]{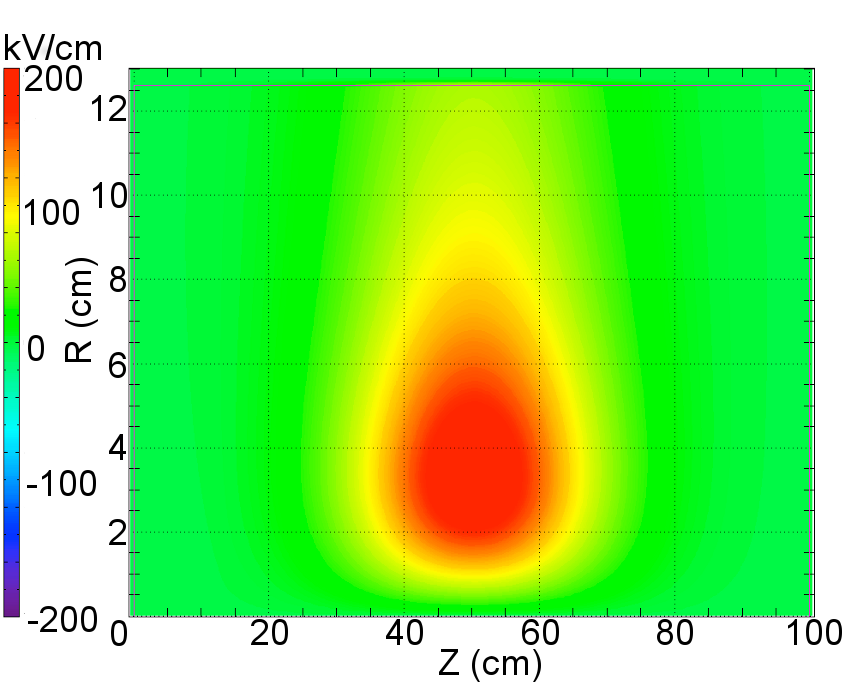} 
(b)\includegraphics[width=0.42\columnwidth]{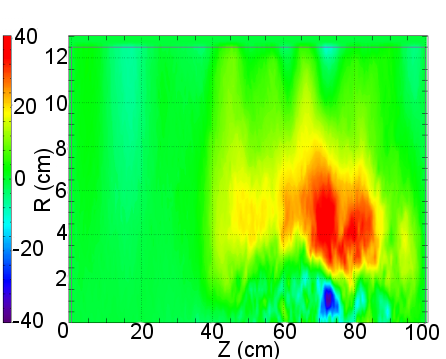} 
(c)\includegraphics[width=0.42\columnwidth]{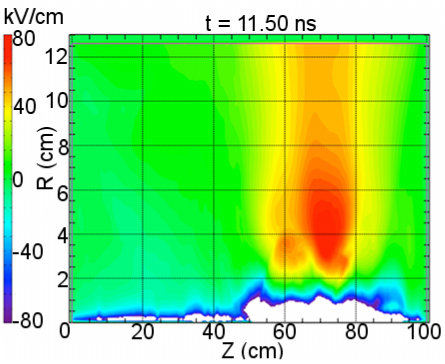} 
(d)\includegraphics[width=0.44\columnwidth]{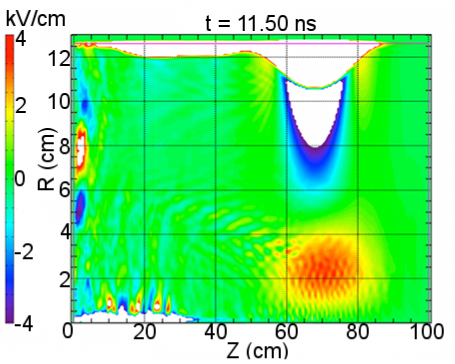} 
\label{fig:quad}
\end{figure}

We used the particle-in-cell code LSP to calculate the effects of the underdense plasma on neutralization. In the simulation, the chamber was of radius 13 cm and length 1 m. We used two-dimensional cylindrical geometry.  The grid spacing was 0.125 cm in both dimensions. The beam density profile was Gaussian in both $r$ and $z$, represented as $n_b = n_0 \exp(r^2/\sigma_r^2+z^2/\sigma_z^2)$, with $\sigma_r=3$ cm and $\sigma_z=25$ cm. The beam velocity in all cases was $\beta=0.34$ and was formed of Pb$^+$ ions, giving a kinetic energy of 12 GeV per ion. These parameters are based on heavy-ion fusion driver scales\cite{roy_drift_2005,friedman_beam_2010,lifschitz_dynamics_2005,vay_intense_2001}. Simulations were carried out in the lab frame. The peak beam density was $n_b = 1.2 \e{11} \mathrm{cm}^{-3}$, and the total beam charge was 3.75 $\mu$C. Background plasma in all cases was completely cold (0 eV). For comparison purposes, the self-fields of the beam propagating through vacuum were $E_r \sim 200$ kV/cm and $B_\theta \sim 200$ G (Fig. \ref{fig:quad}a).  

We consider several schemes of increasing complexity in order to determine the minimum requirements for neutralization. The first scheme we consider is that of electron emitters on the chamber walls and vacuum otherwise. In agreement with earlier studies\cite{lifschitz_dynamics_2005,kaganovich_physics_2010}, the self-electric field is reduced to $\sim40$ kV/cm, only an 80\% reduction from the bare beam case (Fig. \ref{fig:quad}b). Thus, this scheme does not neutralize the pulse effectively. This is because electrons from the walls are accelerated through the large bare beam self-fields, gaining high transverse momentum and oscillating around the beam pulse rather than travelling with it. 

Next, we add background plasma with no electron source on the walls of the chamber. If the plasma has insufficient space-charge, effective neutralization is impossible, evidenced by Fig. \ref{fig:quad}c. However, if there is sufficient space charge (Fig. \ref{fig:quad}d), a high degree of neutralization is achieved. The beam self-electric field is reduced by 98\% from the bare beam case to  $\sim4$ kV/cm. Nonetheless, strong electric fields at the edges of the plasma can be created (such as for $R=8-12 \ \mathrm{cm}$ in Fig. \ref{fig:quad}d), which could damage the focusing chamber.

\begin{figure}[ht!]
\caption{(a) Beam self-electric field after neutralization by volumetric plasma of density $n_p = n_b/5$, with electron emitters on the chamber walls. The self-electric field is reduced by 98.5\%. Contour plot of $E_r$ as a function of $r$ and $z$. (b) Beam self-electric field after neutralization by volumetric plasma of density $n_p = n_b/15$ with an electron source on the chamber walls. The self-electric field is reduced by 98\%. Red represents the analytic model, green represents the LSP simulation with dense plasma on the chamber walls and black represents the LSP simulation with electron emitters on the chamber walls.} 
(a)\includegraphics[width=0.7\columnwidth]{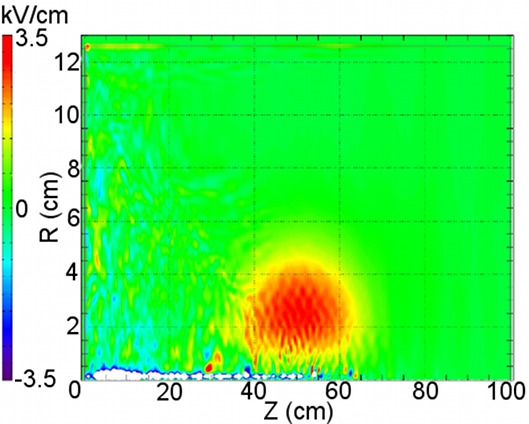} \\
(b)\includegraphics[width=0.7\columnwidth]{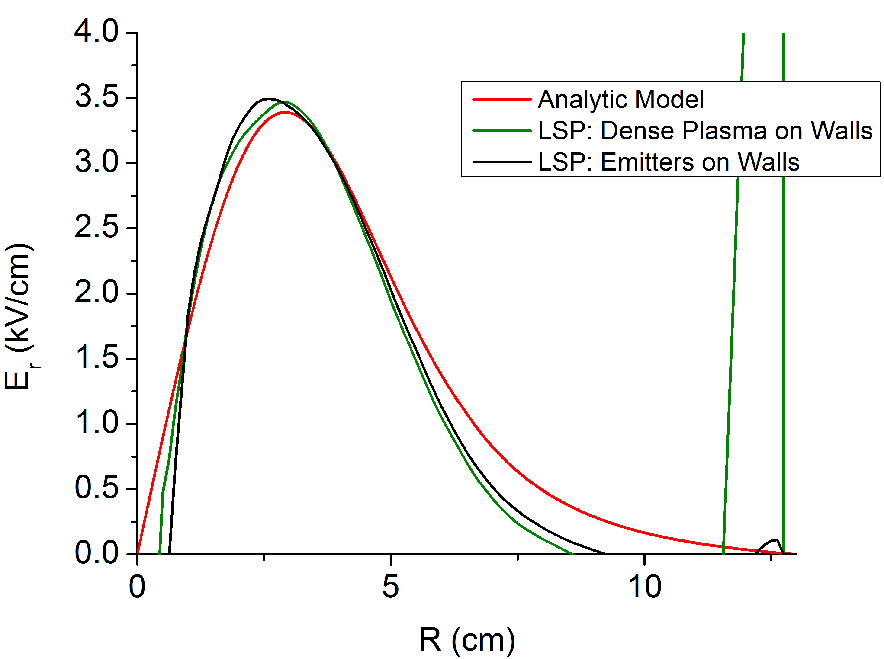}
\label{fig:good_neutralization} 
\end{figure}

Now we combine underdense background plasma with electron emitters on the chamber walls. Here the plasma shorts the strong beam self-fields, allowing electrons to be drawn from the walls without significant transverse heating. As seen in Fig. \ref{fig:good_neutralization}a, the beam self-electric field is reduced by 98.5\% to $\sim 3$ kV/cm without strong electric fields in the plasma. As the beam enters the plasma, electrons are pulled from the plasma and then from the walls of the chamber, leading to neutralization that is initially reliant on hot electrons from the walls. However, electrons are not accelerated in the $z$-direction to the beam's full velocity. This is evidenced by examining the current neutralization. The self-magnetic field is neutralized from the bare case of 200G to 65 G, implying a non-zero net current in the same direction as the beam pulse. Thus, electrons are flushed from the rear of the beam pulse as it travels. The density of the background plasma determines how quickly cold electrons from the plasma are picked up by the beam, replacing the hot electrons from the walls. As the beam picks up cold plasma electrons, the beam self-fields decrease, and hot electrons from the walls can escape the beam potential. In general, the background plasma density determines the distance over which hot electrons are flushed, so there will be a minimum density that can provide effective neutralization for a given chamber length. 

\begin{figure}
\caption{Density slices along $r=2$ cm (to avoid numerical singularities at $r=0$). Black is beam ion density $n_b$, red is emitted electron density $n_e$, dark green is plasma electron density less plasma ion density ${n_p}_e-{n_p}_i$, and bright green is the sum of the red and dark green curves. Here, $n_p=n_b/15$. Hot electrons from the walls are flushed and gradually replaced by cold electrons from the background plasma, producing high levels of neutralization.} 
\includegraphics[width=\columnwidth]{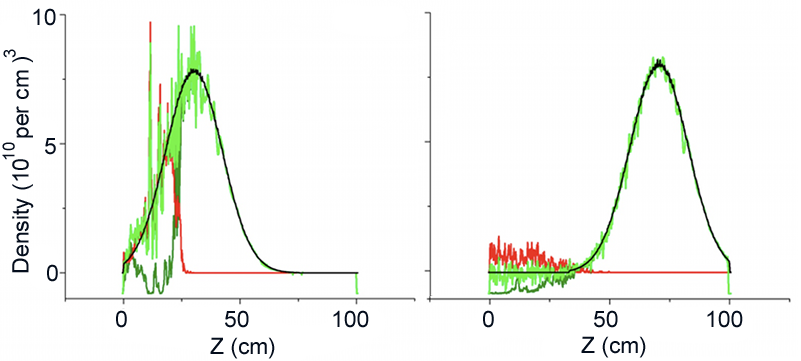}
\label{fig:hot_electrons} 
\end{figure}

Taking a weaker plasma with $n_p = n_b/15$, we again see a 98\% reduction in the beam self-electric field (Fig. \ref{fig:good_neutralization}b). We see that the initial reliance on hot electrons from the walls is strong (Fig. \ref{fig:hot_electrons} left), but that within about 30 cm, these hot electrons are completely replaced by cold plasma (Fig. \ref{fig:hot_electrons} right). A more experimentally viable scheme is to replace the electron emitters by creating a layer of dense plasma near the walls and to fill the chamber with a weak background plasma. This scheme neutralizes the beam just as well as in the case of weak background plasma with electron emitters (Fig. \ref{fig:good_neutralization}b). Over the length of one meter, a plasma of density $n_p = n_b/30$ is not able to achieve high neutralization, as our simulations show that the electric field when the beam exits the chamber is only reduced by 95\% to 10 kV/cm. Longer neutralization chambers will have a lower minimum background plasma density needed for effective neutralization.

We also compare the results of these LSP simulations with an earlier analytic model\cite{kaganovich_nonlinear_2001}. In this model, the authors utilize conservation of generalized vorticity $\Omega = \nabla \times p_e - e B/c$, which has the form $\pd{\Omega}{t} - \nabla \times {V_e} \times {\Omega} = 0$, to find that

\begin{equation}
B = \frac{c}{e} \nabla \times p_e.
\end{equation}  

Coupling this with the electron fluid continuity equation and force balance equation, and assuming a sufficiently long beam, they obtain

\begin{equation}
-\frac{1}{r} \pd{}{r} \left[r\pd{{p_e}_z}{r}\right] = \frac{4\pi e^2}{c^2} (Z_b n_b {V_b}_z - n_e {V_e}_z).
\end{equation}

\begin{figure}
\caption{Comparison of (a) the $E_r$ fields and (b) the $B_\theta$ fields from the analytic model of \cite{kaganovich_nonlinear_2001} and LSP particle-in-cell simulation for the case of $n_p = n_b/5$ with electron emitters on the chamber walls. These are radial slices taken at the center of the beam. Red represents the analytic model, black represents the LSP simulation. There is close agreement.}
\begin{tabular}{ll} 
(a)\includegraphics[width=0.45\columnwidth]{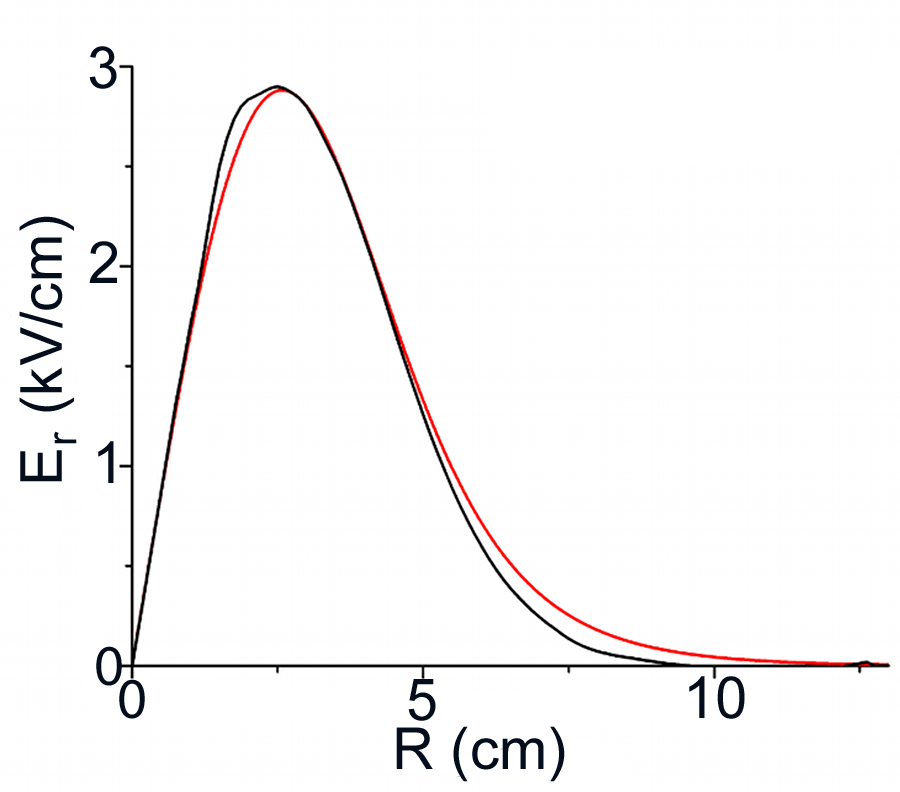} & (b)\includegraphics[width=0.45\columnwidth]{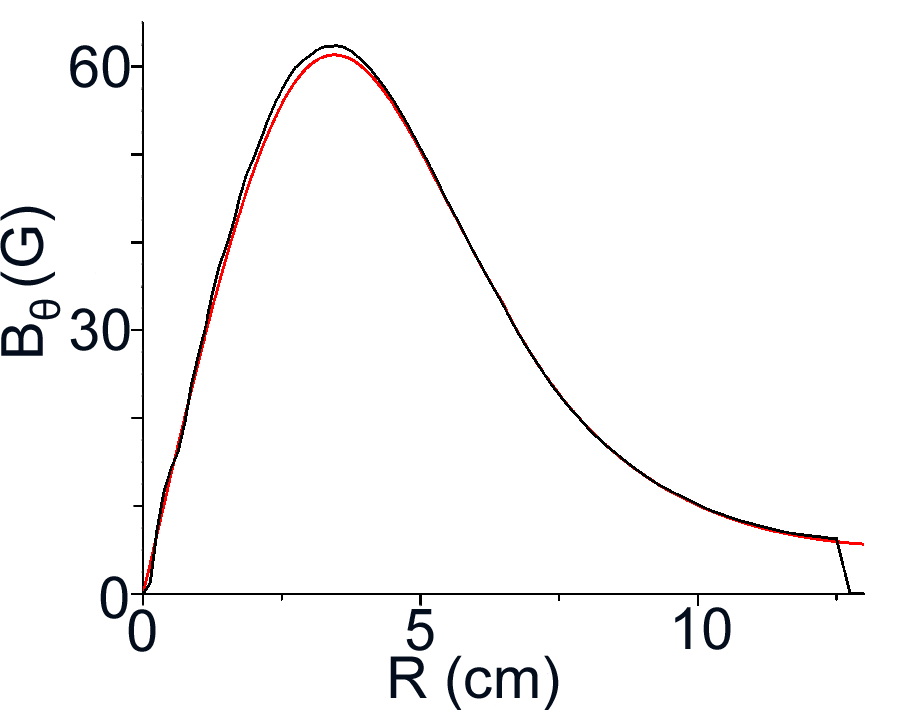}
\end{tabular}
\label{fig:MathCad_vs_LSP}
\end{figure}

Finally, using the continuity equation and the assumption of quasineutrality, the authors obtain an expression for the electric field:

\begin{equation}
E = -\frac{1}{e} \left(V_b \pd{p_e}{\zeta}+\nabla {K_e}\right),
\end{equation}

where $\zeta = V_b t - z$ and $K_e$ is the electron kinetic energy. We evaluate this model for the case of $n_p = n_b/5$ and find that it agrees closely with PIC simulations (Fig. \ref{fig:MathCad_vs_LSP}). 

This effect is also confirmed by experiments on NDCX. In these experiments, a heavy ion beam was propagated through varying densities of underdense plasma with the expectation that the beam radius should diverge as the background plasma becomes less and less dense. This background plasma was created with an FCAPS (Filtered Cathodic Arc Plasma Source) discharge source. The plasma density was inferred from recorded FCAPS discharge voltages, where a larger discharge voltage produced a denser background plasma\cite{roy_space-charge-neutralizing_2009}. To parametrically vary the plasma density upstream of the target and in the final focusing solenoid, the FCAPS discharge voltage was varied. The beam transverse distribution was measured via scintillator at each plasma discharge voltage setting. The bunch compression was also recorded with a fast Faraday cup (FFC). The FCAPS fired reliably from 1 kV to 0.1 kV. A discharge voltage of 0.1 kV created a background plasma of density $3\times10^{11} \ \mathrm{cm}^{-3}$, which is 10 times less than the beam density. Below 0.1 kV, the triggering of the FCAPS was unreliable, with some of the four plasma sources occasionally not firing. For normalization purposes, assuming a mean plasma velocity of $2 \times10^4 \ \mathrm{m/s}$, the four-FCAPS system provided a peak plasma density of $9 \times 10^{12} - 6\times 10^{13}\ \mathrm{cm}^{-3}$ and a 770 A discharge current.

\begin{figure}
\caption{(a) Schematic of experimental setup. (b) Background plasma density as a function of axial position (red) compared with calculated beam density (blue). The beam density was much greater than the plasma density at the target.}
\includegraphics[width=\columnwidth]{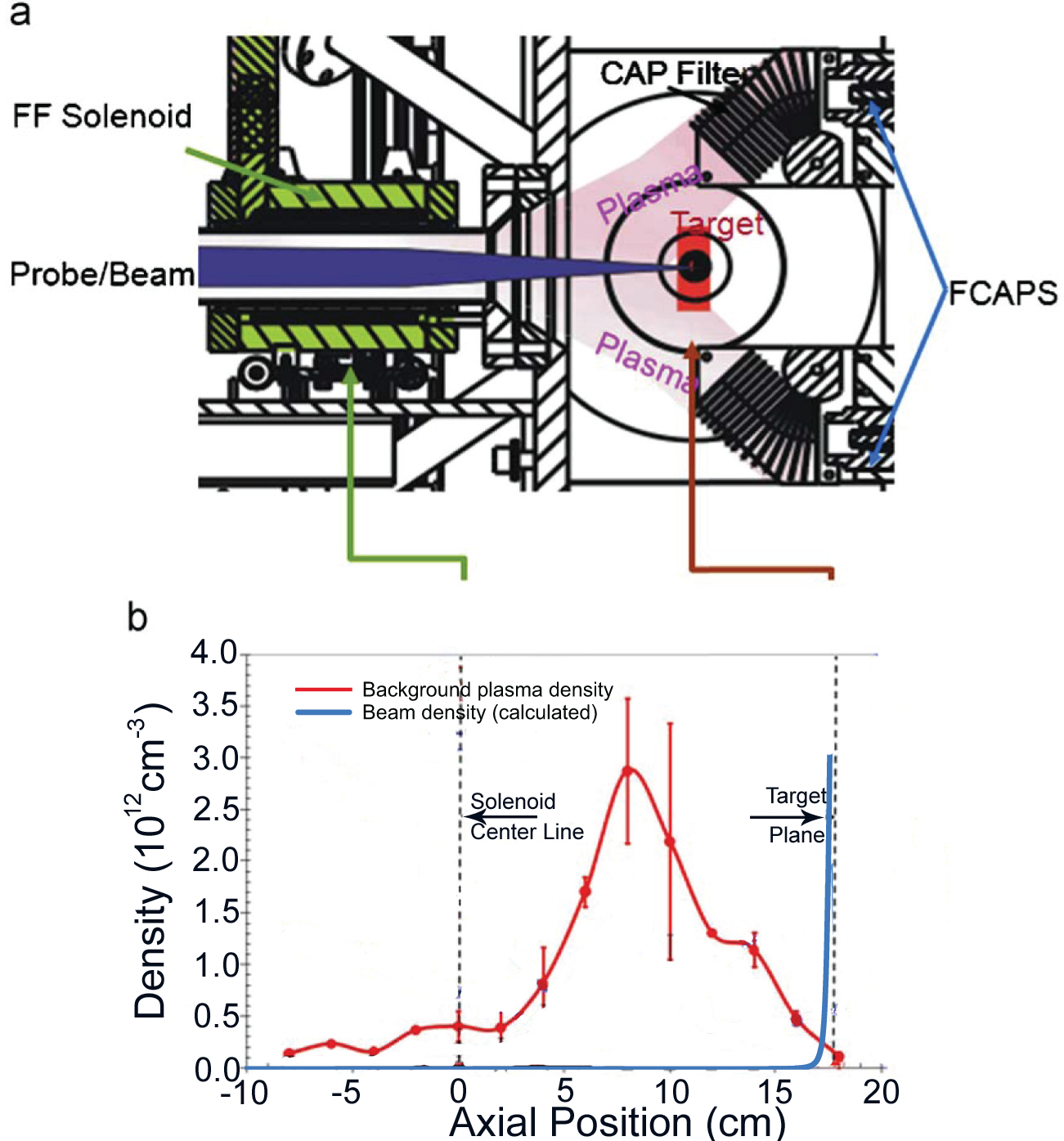} 
\label{fig:Prabir_setup_density} 
\end{figure}

\begin{figure}
\caption{Average of beam envelope parameters $a$ and $b$ ($a = 2 * x_\mathrm{rms}, b = 2 * y_\mathrm{rms}$) for the axially compressed bunch vs background plasma density, inferred from FCAPS discharge voltage. This data analysis was performed by subtracting 392 counts from each of the scintillator images. This subtraction corresponds to approximately a factor of $1/e$ or 0.37 of the peak intensity of a typical scintillator image.} 
\includegraphics[width=0.8\columnwidth]{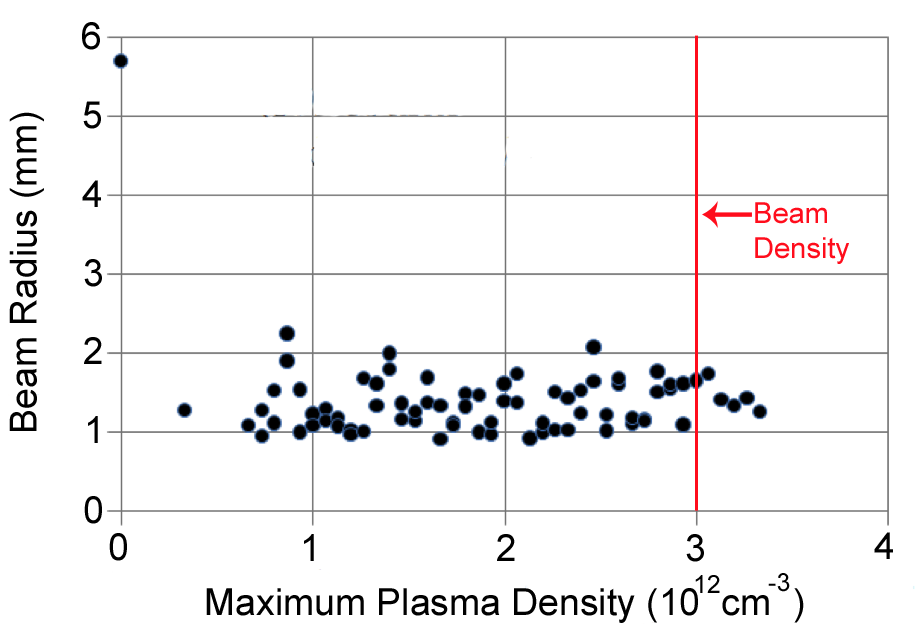}
\label{fig:Prabir_radius_vs_density}  
\end{figure}

The beam radius and peak intensity did not change significantly with the variation of background plasma density over the range of reliable FCAPS firing, where the background plasma density varied from $5 \times 10^{11}\ \mathrm{cm}^3$ to $3.3 \times 10^{12} \ \mathrm{cm}^3$, as shown in Figures \ref{fig:Prabir_setup_density} and \ref{fig:Prabir_radius_vs_density}. The data in Fig. \ref{fig:Prabir_radius_vs_density} was taken using scintillator image analysis with background subtraction. In the case of no background plasma, the beam diverged to a final radius of 5.75 cm, 2.3 times its initial radius of 2.5 cm. However, the beam converged to a radius of around 1 cm for the range of background plasma densities $n_p = n_b/6$ to $n_p = n_b$. Thus, weak background plasma near the target effectively neutralized the beam's self-electric field. 

In conclusion, we have shown that a high degree of neutralization can be achieved by propagating an ion beam pulse through underdense background plasma with either a large amount of space charge or an electron source on the walls of the neutralization chamber. The beam self-electric field is reduced by 98\%, high enough for inertial confinement fusion applications. There are many disadvantages to using a dense background plasma to neutralize intense ion beam pulses, chiefly that dense plasmas are difficult to produce and require a large amount of energy; we have shown that a weak background plasma may suffice under certain conditions.

\bibliography{NeutralizationPaperBibliography}
\bibliographystyle{apsrev4-1}

\end{document}